\documentclass[preprint,showpacs,aps,amssymb,floatfix,prd,amsmath,preprintnumbers,showkeys]{revtex4} 
\usepackage{epstopdf}
\usepackage{capt-of}
\usepackage{graphicx}  
\usepackage{dcolumn}   
\usepackage{bm}
\begin{document}
\input epsf.tex

\title{Energy and momentum of  Bianchi Type $VI_h$ Universes\footnote{\scriptsize A part of this collaborative research  work was done during an International  Workshop on  Introduction to Research in Einstein's General Relativity at NIT, Patna (India). Authors' emails as well as their  permanent addresses are mentioned below:}}

\author	{
	S. K. Tripathy\footnote{\scriptsize {Department of Physics, Indira Gandhi Institute of Technology, Sarang, Dhenkanal, Odisha-759146, India, Email: tripathy\_sunil@rediffmail.com}},
	B. Mishra\footnote{\scriptsize {Department of Mathematics, Birla Institute of Technology and Science-Pilani, Hyderabad Campus, Hyderabad-500078, India, Email:	bivudutta@yahoo.com}},
	G. K. Pandey\footnote{\scriptsize{Patna Science College, Patna University, Patna 80005, Bihar, India, Email: gaurav.golu1@gmail.com}},	
	A. K. Singh\footnote{\scriptsize {Physics department, Patna University, Patna 80005, Bihar, India, Email:         singh.abhaykant@gmail.com}},	
	T. Kumar\footnote{\scriptsize { Physics department, Patna University, Patna 80005, Bihar, India, 
               Email: tinkujaymatadi@gmail.com}}}
\affiliation{National Institute of Technology, Patna 800005, Bihar, India. \\  }

\author {S.~S.~Xulu\footnote{\scriptsize{ Email: ssxulu@pan.uzulu.ac.za}}}
    \affiliation{Department of Computer Science, University of Zululand,3886 Kwa-Dlangezwa, South Africa. \vspace*{0.5in}}


\begin{abstract}

We obtain the energy and momentum of the   Bianchi type $VI_h$  universes using different prescriptions for the energy-momentum complexes in the framework of general relativity.  The energy and momentum of the Bianchi $VI_h$ universse  are found to be zero  for the  parameter $h = -1$ of the metric. The vanishing of these results support the  conjecture of Tryon that 
Universe must have a zero net value for all conserved  quantities.This also supports the work of Nathan Rosen  with the Robertson-Walker metric. Moreover, it raises an interesting question: ``Why $h=-1$ case is so special?"

\end{abstract}

\pacs{04.20.Jb, 04.20.Cv}


\keywords{Bianchi Type $VI_h$ metric, Energy-momentum complex, General Relativity.}

\maketitle


\section{Introduction}
The local distribution of energy and momentum has remained a challenging  domain of research in the context of Einstein's general relativity.  Einstein proposed the first energy-momentum complex \cite{Ein15} that follows the covariant conservation laws by including the energy and momenta of gravitational fields along with those of matter and non-gravitational fields. The energy-momentum due gravitational field turns out to be a non-tensorial object. The choice of gravitational field pseudotensor (non-tensor) is not unique and therefore, following Einstein, many authors prescribed different definitions of energy-momentum complexes based on canonical approach (e.g. Tolman \cite{Tol34}, Papapetrou \cite{Pap48} , Landau and Lifshitz(LL) \cite{LL51} , Bergmann and Thomson (BT) \cite{Berg53}, and   Weinberg \cite{Wein72}.)  The Tolman definition is essentially the same as that of Einstein; however, these two definitions differ in form and sometimes it is easier to use Tolman's definition. This was explained  by Virbhadra\cite{Vir1999}. The main concern in the use of these definitions is that they are coordinate-dependent. However, with these definitions, meaningful and reasonable results can be obtained if ``Cartesian coordinates" (also called quasi-Cartesian or  quasi-Minkowskian for asymptotically Minkowskian space-times)  are used.  Some coordinate-independent definitions have also been proposed by  M{\o}ller \cite{Moll58} , Komar \cite{Komar}  and Penrose \cite{Pen82} . The coordinate-independent prescriptions, including   quasi-local mass of Penrose \cite{Pen82}  were found to have some serious shortcomings as these are  limited to certain class of symmetries only (see in \cite{Vir1999} and also references therein.)  

The issue of energy localization and the coordinate dependence of these definitions gained momentum with renewed interests after the works of Virbhadra and his collaborators (notably, Nathan Rosen - the most famous collaborator of Albert Einstein - of  Einstein-Rosen bridge, the EPR paradox, and Einstein-Rosen gravitational waves fame) found a striking similarity in the results for different energy momentum prescriptions. 
They considered numerous space-times\cite{Rose,KSV} and obtained seminal results that rejuvenated this field  of research.

Virbhadra \cite{Vir1999} further investigated  whether or not these energy-momentum complexes lead to the same results for the most general non-static spherically symmetric metric and found that they disagree. He and his collaborators\cite{KSV} observed that if the calculations of energy momentum  are done in Kerr-Schild Cartesian  coordinates, then the energy-momentum complexes of  Papapetrou \cite{Pap48}, Landau and Lifshitz \cite{LL51}, and Weinberg \cite{Wein72} produce the same result as in Einstein definition. However, if the computation are made in Schwarzschild Cartesian coordinates, these energy momentum complexes  disagree \cite{Vir1999}. Xulu \cite{Xulu03} confirmed this by obtaining the energy and momentum for most general non-static spherically symmetric system using  M{\o}ller definition and found different result in general form than those obtained in Einstein's definition.  Xulu and  others \cite{XuluAndMany} obtained many  important results in this field.
However,  till now, there is no completely acceptable definition for energy and  momentum distributions in general relativity even though prescriptions in teleparallel gravity claim to provide satisfactory solution to this problem \cite{Sharif10}. 

Bianchi type models are spatially homogeneous and anisotropic universe models. These models are nine in number, but their classification permits to split them in two classes. There are six models in class A ($I$, $II$, $VI_1$, $VII$, $VIII$ and $IX$) and five in class B ($III$, $IV$, $V$, $VI_h$ and $VII_h$). Spatially homogeneous cosmological models play an important role to understand the structure and properties of the space of all cosmological solutions of Einstein field equations.  These spatially homogeneous and anisotropic models are the exact solutions of Einstein Field equations and are more general than the Friedman models in the sense that they can provide interesting results pertaining to the anisotropy of the universe. It is worth to mention here that, the issue of global anisotropy has gained a lot of research interest in recent times. The standard cosmological model ($\Lambda$CDM) based upon the spatial isotropy and flatness of the universe is consistent with the data from  precise measurements of the CMB temperature anisotropy \cite{Hinshaw09} from Wilkinson Microwave Anisotropy Probe(WMAP). However, the $\Lambda$CDM model suffers from some anomalous features at large scale and signals a deviation from the usual geometry of the universe. Recently released Planck data \cite{Ade} show a slight red-shift of the primordial power spectrum from the exact scale invariance. It is clear from the Planck data that, $\Lambda$CDM model does not fit well to the temperature power spectrum at low multipoles. Also, precise measurements from WMAP predict asymmetric expansion with one direction expanding differently from the other two transverse directions at equatorial plane \cite{Buiny06} which signals a non trivial topology of the large scale geometry of the universe (\cite{Wata09, SKT14} and references therein).
Using Einstein definition, Banerjee and Sen \cite{Ban97} studied  energy distribution with Bianchi type I (BI) space-time. 

In recent times this pressing issue of energy and momentum localization has been studied widely by many authors using different space-times and definitions of energy momentum complexes. The importance of the study of energy and momentum distribution lies in the fact that it helps us  getting an idea of effective  gravitational mass of  metrics of certain symmetries and can put deep insight into the gravitational lensing phenomena\cite{Lens}.

Xulu \cite{Xulu2000} calculated the total energy in BI universes using the prescriptions of LL, Pappapetrou and Weinberg. Radinschi \cite{Rad01} calculated the energy of a model of the universe based on the Bianchi type $VI_0$ metric using the energy-momentum complexes of LL and of Papapetrou. She found that the energy due to the matter plus field is equal to zero.  Aydogdu and Salti \cite{AS05}, using the M{\o}ller's tetrad investigated the energy of BI universe. In another work,  Aydogdu and Salti \cite{AS06}, calculated the energy of LRS Bianchi type II metric to get consistent result.

In this paper, we obtain the energy and momentum for a more general homogeneous and anisotropic Bianchi type $VI_h$ metric and its transformation by using different prescriptions for the energy-momentum complex in general relativity. Bianchi type $VI_h$ model have already been shown to provide interesting results in cosmology in connection with the late time accelerated expansion of the universe when the contribution to the matter field comes from one dimensional cosmic strings and bulk viscosity \cite{SKT10}. In this work, we have used the convention that Latin indices take value from 0 to 3 and Greek indices run from 1 to 3. We use geometrized units where $G=1$ and $c=1$. The composition of the paper is as follows. In section 2, we have write the energy momentum tensor for Bianchi type $VI_h$ space-time. In section 3,  the Einstein energy momentum complex is discussed, following which we investigate the energy momentum complex of Landau and Lifshitz, Papapetrou and Bergmann Thomson for the assumed metric in the subsequent subsections. In the last section, we summarize our results.


\section{Bianchi type $VI_h$ space-time}
We have considered Bianchi type $VI_h$ space-time in the form

\begin{equation}
ds^{2}=-dt^{2}+A^{2}dx^{2}+B^{2}e^{2x}dy^{2}+C^{2}e^{2hx}dz^{2}.\label{eq:1}
\end{equation}
The metric potentials $A$, $B$ and $C$ are functions of cosmic time $t$ only. Further, $x^i,i=1,2,3,0$ respectively denote for the coordinates $x$,$y$,$z$ and $t$ and the exponent $h$ can assume integral values in the range   $h=-1,0,1$.

For the metric  $\eqref{eq:1}$ , the determinant of the metric tensor $g$ and the contravariant components of the tensor are respectively given as:

\begin{eqnarray}
g &=& |g_{ab}| = -A^2B^2C^2 e^{2(h+1)x}\text{,}\nonumber \\
g^{00} &=&-1\text{,}\nonumber \\
g^{11} &=&\frac{1}{A^2}\text{,}\nonumber \\
g^{22} &=&\frac{1}{B^2 e^{2x}}\text{,}\nonumber \\
g^{33} &=&\frac{1}{C^2 e^{2hx}}.\label{eq:2}
\end{eqnarray}
In General Relativity, the energy-momentum tensor is given by $8 \pi T_i^j=R_i^j-\frac{1}{2}Rg_i^j$, where $R_i^j$ is the Ricci tensor and $R$ is the Ricci scalar. The non-vanishing components of the energy-momentum tensor for the Bianchi type $VI_h$ space time is given below (this is  not a new result; however, we write here  because we need these for  analysing and discussing results.)

\begin{eqnarray}
8 \pi T_1^1 &=& \frac {\ddot{B}}{B}+\frac
  {\ddot{C}}{C}+\frac {\dot{B} \dot{C}}{BC} -\frac {h}{A^2},\label{eq:3}\\
8 \pi  T_2^2 &=& \frac {\ddot{A}}{A}+\frac
   {\ddot{C}}{C}+\frac {\dot{A}\dot{C}}{AC} -\frac {h^2}{A^2}, \label{eq:4}\\
8 \pi T_3^3 &=& \frac {\ddot{A}}{A}+\frac
  {\ddot{B}}{B}+\frac {\dot{A}\dot{B}}{AB} -\frac {1}{A^2},\label{eq:5}\\
8 \pi  T_0^0 &=& -\frac {\dot{A}\dot{B}}{AB}-\frac {\dot{B}\dot{C}}{BC}-
   \frac {\dot{A}\dot{C}}{AC} +\frac{1+h+h^2}{A^2},\label{eq:6}\\
8 \pi  T_0^1 &=& (1+h)\frac {\dot{A}}{A}-\frac {\dot{B}}{B}
  -h\frac {\dot{C}}{C},\label{eq:7}
\end{eqnarray}
where the overhead dots hereafter, denote ordinary time derivatives.
 

\section{Energy-Momentum Complexs}

\subsection{Einstein Energy-momentum Complex}

The Einstein energy-momentum complex is \cite{Ein15}

\begin{equation}
\Theta_i^k=\frac{1}{16\pi}H_{i,l}^{kl},\label{eq:8}
\end{equation}
where
\begin{equation}
H_i^{kl}=-H_i^{lk}=\frac{g_{in}}{\sqrt{-g}}[-g(g^{kn}g^{lm}-g^{ln}g^{km})]_{,m}.\label{eq:9}
\end{equation}
$\Theta_0^0$ and $\Theta_\alpha^0$ stand for the energy and momentum density components respectively. The energy and momentum components are obtained through a volume  integration 

\begin{equation}
P_i=\int\int\int \Theta_i^0 dx^1dx^2dx^3.\label{eq:10}
\end{equation}

By applying Gauss's  theorem, the above equation can also be reduced to
\begin{equation}
P_i=\frac{1}{16\pi}\int\int H_i^{0\alpha} n_\alpha dS, \label{eq:11}
\end{equation}
where $n_\alpha$ is the outward unit normal vector over the infinitesimal surface element $dS$. $P_0$ and $P_\alpha$ stand for the energy and momentum components respectively.
The required non-vanishing components of the $ H_i^{kl}$ for the line element $\eqref{eq:1}$ are given by
\begin{eqnarray}
H_0^{01} &=& -\frac{2BC}{A}(1+h)e^{x(1+h)},\nonumber\\
H_1^{01} &=& -2ABC \left(\frac{\dot{B}}{B}+\frac{\dot{C}}{C}\right)e^{x(1+h)},\nonumber\\
H_2^{02} &=& -2ABC \left(\frac{\dot{A}}{A}+\frac{\dot{C}}{C}\right)e^{x(1+h)},\nonumber\\
H_3^{03} &=& -2ABC \left(\frac{\dot{A}}{A}+\frac{\dot{B}}{B}\right)e^{x(1+h)}.\label{eq:12} 
\end{eqnarray}
Using $\eqref{eq:12}$, we obtain the components of energy and momentum densities as
\begin{eqnarray}
\Theta_0^0 &=& -\frac{BC}{8\pi A}\left(1+h\right)^2 e^{x(1+h)},\nonumber\\
\Theta_1^0 &=& -\frac{ABC}{8\pi}\left(\frac{\dot{B}}{B}+\frac{\dot{C}}{C}\right)(1+h) e^{x(1+h)},\nonumber\\
\Theta_2^0 &=& \Theta_3^0=0.\label{eq:13}
\end{eqnarray}
If the dependence of $A, B, C$ and $h$ on the time coordinate $t$ were known, one could  evalate the surface intergral. Even in the absence of such  results, it is clear thaat the energy of the $VI_h$ universe in Einstein prescription is  not zero for $h=0,1$. However, it is interesting to note  that, the energy and momentum densities vanish for $h=-1$.
\subsection{Landau and Lifshitz Energy-momentum Complex}

The symmetric Landau and Lifshitz energy-momentum complex is  \cite{LL51}
\begin{equation}
L^{ik}=\frac{1}{16\pi}\lambda_{,lm}^{iklm},\label{eq:14}
\end{equation}
where 

\begin{equation}
\lambda^{iklm}=-g(g^{ik}g^{lm}-g^{il}g^{km}).\label{eq:15}
\end{equation}
Here $L^{00}$ and $L^{\alpha 0}$ are the energy and  energy density components.

The energy and momentum can be defined  through the volume integral
\begin{equation}
P^i=\int\int\int L^{i0}dx^1dx^2dx^3.\label{eq:16}
\end{equation}
Using Gauss theorem, the energy and momentum components become
\begin{equation}
P^i=\frac{1}{16\pi}\int\int \lambda ^{i0\alpha m}_{,m} n_{\alpha} dS.\label{eq:17}
\end{equation}

The required non-vanishing components of $\lambda^{iklm}$ for the present model are obtained as

\begin{eqnarray}
\lambda^{0011} &=& -B^2C^2e^{2x(1+h)},\nonumber\\
\lambda^{1010} &=& B^2C^2e^{2x(1+h)},\nonumber\\
\lambda^{0022} &=& -A^2C^2e^{2hx},\nonumber\\
\lambda^{2020} &=& A^2C^2e^{2hx},\nonumber\\
\lambda^{0033} &=& -A^2B^2e^{2x},\nonumber\\
\lambda^{3030} &=& A^2B^2e^{2x}.\label{eq:18}
\end{eqnarray}

Using these components in $\eqref{eq:17}$, we obtain

\begin{eqnarray}
P^0 &=& -\frac{xr}{2}(1+h)B^2C^2e^{2x(1+h)},\nonumber\\
P^1 &=& \frac{xr}{2}\left(\frac{\dot{B}}{B}+\frac{\dot{C}}{C}\right)B^2C^2e^{2x(1+h)},\nonumber\\
P^2 &=& \frac{yr}{2}\left(\frac{\dot{A}}{A}+\frac{\dot{C}}{C}\right)A^2C^2e^{2hx},\nonumber\\
P^3 &=& \frac{zr}{2}\left(\frac{\dot{A}}{A}+\frac{\dot{B}}{B}\right)A^2B^2e^{2x},\label{eq:19}
\end{eqnarray}
where $r=\sqrt{x^2+y^2+z^2}$. For Landau and Lifshitz prescription, the energy of the universe is non zero for $h=0$ and $1$ and it vanishes for $h=-1$.
\subsection{Papapetrou Energy-momentum Complex}

The symmetric energy-momentum complex of Papapetrou  is given by \cite{Pap48}
\begin{equation}
\Sigma^{ik}=\frac{1}{16\pi} {\mathcal{N}}^{iklm}_{,lm},\label{eq:20}
\end{equation}
where

\begin{equation}
{\mathcal{N}}^{iklm}=\sqrt{-g}(g^{ik}\eta^{lm}-g^{il}\eta^{km}+g^{lm}\eta^{ik}-g^{lk}\eta^{im}),\label{eq:21}
\end{equation}
with
\begin{equation}
\eta^{ik}=diag(-1,1,1,1).\nonumber\\
\end{equation}
Here $\Sigma^{00}$ and $\Sigma^{\alpha 0}$ are respectively the energy and energy current (momentum) density components. The energy and momentum can be defined as
\begin{equation}
P^i=\int\int\int \Sigma^{i0}dx^1dx^2dx^3.\label{eq:22}
\end{equation}
For time independent metrics, one can have
\begin{equation}
P^i=\frac{1}{16\pi}\int\int {\mathcal{N}}^{i0\alpha \beta}_{,\beta} n_{\alpha} dS.\label{eq:23}
\end{equation}
The  non-vanishing components of ${\mathcal{N}}^{iklm}$ required to obtain the energy and momentum density components for the space-time described by the line element $\eqref{eq:1}$ are

\begin{eqnarray}
{\mathcal{N}}^{1001} &=& {\mathcal{N}}^{2002}={\mathcal{N}}^{3003}=ABC e^{x(1+h)},\nonumber\\
{\mathcal{N}}^{0011} &=& -ABC e^{x(1+h)}\left(1+\frac{1}{A^2}\right),\nonumber\\
{\mathcal{N}}^{1010} &=& ABC e^{x(1+h)}\left(\frac{1}{A^2}\right),\nonumber\\
{\mathcal{N}}^{0022} &=& -ABC e^{x(1+h)}\left(1+\frac{1}{B^2e^{2x}}\right),\nonumber\\
{\mathcal{N}}^{2020} &=& ABC e^{x(1+h)}\left(\frac{1}{B^2e^{2x}}\right),\nonumber\\
{\mathcal{N}}^{0033} &=& -ABC e^{x(1+h)}\left(1+\frac{1}{C^2e^{2hx}}\right),\nonumber\\
{\mathcal{N}}^{3030} &=& ABC e^{x(1+h)}\left(\frac{1}{C^2e^{2hx}}\right).\label{eq:24}
\end{eqnarray}
The Papapertrou energy and energy current density components are obtained by using the above components in \eqref{eq:24} as 

\begin{eqnarray}
\Sigma^{00} &=& -\frac{(1+h)^2}{16\pi} ABC \left(1+\frac{1}{A^2}\right)e^{x(1+h)},\nonumber\\
\Sigma^{10} &=& \frac{1+h}{16\pi}\left[\left(\frac{\dot{A}}{A}+\frac{\dot{B}}{B}+\frac{\dot{C}}{C}\right)+\frac{1}{A^2}\left(\frac{\dot{B}}{B}+\frac{\dot{C}}{C}-\frac{\dot{A}}{A}\right)\right] ABC  e^{x(1+h)},\nonumber\\
\Sigma^{20} &=& 0, \nonumber\\
\Sigma^{30} &=& 0. \label{eq:32}
\end{eqnarray}


Like the previous cases, it can be concluded from the above components that, the energy of the universe in the Papapetrou prescription is non-zero for $h=0$ and $1$ and it becomes zero for $h=-1$. Therefore, energy as well as momentum componnets both are zero for $h=-1$.
\subsection{Bergmann-Thompson Energy-momentum Complex}

The Bergmann-Thompson energy-momentum complex is  \cite{Berg53}
\begin{equation}
{\bf{B}}^{ik}=\frac{1}{16\pi} [g^{il}{\mathcal{B}}^{km}_l]_{,m} ,\label{eq:25}
\end{equation}
where 
\begin{equation}
 {\mathcal{B}}_l^{km}=\frac{g_{ln}}{\sqrt{-g}} [-g(g^{kn}g^{mp}-g^{mn}g^{kp})]_{,p}.\label{eq:26}
\end{equation}
Here ${\bf{B}}^{00}$ and ${\bf{B}}^{\alpha 0}$ are the energy and momentum densities respectively. The energy and momentum can be defined as

\begin{equation}
P^i=\int\int\int {\bf{B}}^{i0}dx^1dx^2dx^3.\label{eq:27}
\end{equation}
Using Gauss theorem, the energy and momentum components can be expressed as
\begin{equation}
P^i=\frac{1}{16\pi}\int\int {\mathcal{B}} ^{i0\alpha} n_{\alpha} dS.\label{eq:28}
\end{equation}

In order to calculate the energy and momentum distribution for Bianchi type $VI_h$ space-time using Bergmann and Thompson energy momentum complex, we require the following non-vanishing components of $B^{km}_l$
\begin{eqnarray}
{\mathcal{B}}_1^{01} &=&  2ABC e^{x(1+h)}\left(\frac{\dot{B}}{B}+\frac{\dot{C}}{C}\right),\nonumber\\
{\mathcal{B}}_2^{02} &=& 2ABC e^{x(1+h)}\left(\frac{\dot{A}}{A}+\frac{\dot{C}}{C}\right),\nonumber\\ 
{\mathcal{B}}_3^{03} &=& 2ABC e^{x(1+h)}\left(\frac{\dot{A}}{A}+\frac{\dot{B}}{B}\right).\label{eq:29}
\end{eqnarray}
The energy and momentum density components can be obtained by using $\eqref{eq:29}$ in $\eqref{eq:25}$ as,

\begin{eqnarray}
{\bf{B}}^{00} &=& {\bf{B}}^{20}={\bf{B}}^{30}=0,\\ \label{eq:30}
{\bf{B}}^{10} &=& \left(\frac{1+h}{8\pi}\right)\frac{BCe^{x(1+h)}}{A}\left(\frac{\dot{B}}{B}+\frac{\dot{C}}{C}\right).\label{eq:31}
\end{eqnarray}

It is clear that, the energy of the Bianchi $VI_h$ universe is zero in Bergmann-Thompson prescription $\forall h = 0, \pm 1$. However, the energy as well as all momentum components vanish for $h=-1$

\section{Summary}

In the present study, we have obtained energy and momentum distributions for spatially homogeneous and anisotropic Bianchi type $VI_h$ metric using Einstein,  Landau and Lifshitz, Papapetrou, and Bergmann-Thompson complexes in the framework of general relativity. Bianchi type $VI_h$ space-time has an edge over the usual Friedman-Robertson-walker (FRW) in the sense that it can handle the anisotropic spatial expansion. In the present study, we found the  energy and momentum vanish for $h=-1$ Bianchi $VI_h$ Universe. The  only exception to this is the case of Landau and Lifshitz where energy density components though vanish, momentum density components do not vanish in general for  $h=-1$ .  Virbhadra (refer to \cite{Vir1999} and references in this paper) however pointed out that the Landau and Lifshitz complex  does not work as good  as Einstein complex and the latter is the best for energy-momentum  calculations. Other energy-momentum complexes agree with that of the Einstein's for $h=-1$ case. Equations $(3-7)$ show that energy and mometum densities due to matter and non-gravitational fields are non-zero even for $h=-1$. However, as energy-momentum complexes include effects of gravitational field as well, the total comes to be zero.

The observation of the cosmic microwave background radiation by Penzias and Wilson\cite{Pen65a,Pen65b} in the year 1965  strongly supports that some version of the big bang theory is correct and it also suggested a  remarkable conjecture regarding the total energy  and momentum of the universe.
Tryon\cite{Tryon} assumed  that our Universe appeared from nowhere about $10^{10}$ years ago. He pondered that the conventional laws of physics need
not have been violated at the time of creation of the Universe. Therefore, he proposed that our Universe must have a zero net value for all conserved quantities. Further,  he presented arguments suggesting that  the net energy and momentum of our Universe may be indeed zero.
Tryon gave a  big bang model and according to that   our Universe is a fluctuation of the vacuum  and he predicted a homogeneous, isotropic and closed Universe consisting of equal amount of   matter and anti-matter.  Tryon, in the same paper,  also mentioned  an excellent  topological argument by Bergmann that any closed universe has zero energy. Later,  Rosen\cite{Rose94}, in the year 1994, considered a  closed homogeneous isotropic universe described by the Friedman-Robertson-Walker (FRW) metric and found that  the total energy  of the Universe zero. (Sadly, he passed away in the following year 1995.)  Excited by these results, Xulu\cite{Xulu2000}  studied energy and momentum in Bianchi type I universes and his results supported the conjecture of Tryon.

In view of the excellent results mentioned in above paragraph, the outcome of our  research that the energy and momentum of $h=-1$ Bianchi $VI_h$ Universe are zero is indeed a very important result. However, one would ask: ``Why is this true for  $h=-1$ case only?"  History of science has  records that coincidence of results usually  point out something very important. Thus, it remains to investigated: ``Why $h=-1$  case is so special?" It is likely that outcome of these investigations would give important implications for general relativity and relativistic astrophysics.

\newpage

\section{Acknowledgments}
Our mentor, Dr. K. S. Virbhadra, cannot be thanked enough for his guidance and incredible  support throughout the preparation of this project. The authors would like to thank  National Institute of Technology (NIT), Patna, India for inviting us to attend the {\em Workshop on Introduction to Research in Einstein's General Relativity} under TEQIP programme held at NIT, Patna during which this collaborative research work was initiated. Also, the authors acknowledge the warm hospitality extended by the Director, Prof. A. De, NIT, Patna and the Convener of the workshop Prof. B. K. Sharma. SKT and BM like to thank Inter University Centre for Astronomy and Astrophysics (IUCAA), Pune, India for providing facility and support during an academic visit where a part of this work was carried out. SSX thanks the University of Zululand (South Africa) for all support. GKP thanks his friends for helping him in checking the calculations.


\end{document}